\documentclass[aps,prd,twocolumn,groupedaddress,showpacs]{revtex4}

\usepackage{graphicx}

\begin{document}


\title{New Constraints on Radiative Decay of Long-Lived Particles in Big Bang 
Nucleosynthesis with New $^4$He Photodisintegration Data}


\author{Motohiko Kusakabe$^{1,2}$\footnote{Research Fellow of the Japan 
Society for the Promotion of Science}\footnote{kusakabe@th.nao.ac.jp}, Toshitaka Kajino$^{1,2,3}$, Takashi Yoshida$^{1}$, Tatsushi Shima$^{4}$,\\
Yasuki Nagai$^{4}$ and Toshiteru Kii$^{5}$}

\affiliation{
$^1$Division of Theoretical Astronomy, National Astronomical Observatory 
of Japan, Mitaka, Tokyo 181-8588, Japan \\
$^2$Department of Astronomy, Graduate School of Science, University of 
Tokyo,  Hongo, Bunkyo-ku, Tokyo 113-0033, Japan \\
$^3$Department of Astronomical Science, The Graduate University for 
Advanced Studies, Mitaka, Tokyo 181-8588, Japan \\
$^4$ Research Center for Nuclear Physics, Osaka University, Ibaraki, 
Osaka 567-0047, Japan \\
$^5$ Institute of Advanced Energy, Kyoto University, Gokasyo, Uji, Kyoto 
611-0011, Japan}


\date{\today}

\begin{abstract}

A recent measurement of $^4$He photodisintegration reactions, 
$^4$He($\gamma$,$p$)$^3$H and $^4$He($\gamma$,$n$)$^3$He with 
laser-Compton photons shows smaller cross sections than those estimated by 
other previous experiments at $E_\gamma \lesssim 30$~MeV. We study 
big-bang nucleosynthesis with the radiative particle decay using the 
new photodisintegration cross sections of $^4$He as well as previous data.  The sensitivity of 
the yields of all light elements D, T, $^3$He, $^4$He, $^6$Li, $^7$Li 
and $^7$Be to the cross sections is investigated.  The change of the 
cross sections has an influence on the non-thermal yields of D, $^3$He and 
$^4$He.  On the other hand, the non-thermal $^6$Li 
production is not sensitive to the change of the cross sections at this low 
energy, since the non-thermal secondary synthesis of $^6$Li needs 
energetic photons of $E_\gamma \gtrsim 50$~MeV. 
The non-thermal nucleosynthesis triggered by the radiative particle 
decay is one of candidates of the production mechanism of $^6$Li 
observed in metal-poor halo stars (MPHSs). In the parameter region of 
the radiative particle lifetime and the emitted photon energy which satisfies 
the $^6$Li production above the abundance level observed in MPHSs, 
the change of the photodisintegration cross sections at $E_\gamma \lesssim 
30$~MeV as measured in the recent experiment leads to $\sim 10$~\%
  reduction of resulting $^3$He abundance, whereas the $^6$Li abundance does not change 
for this change of the cross sections of $^4$He($\gamma$,$p$)$^3$H and 
$^4$He($\gamma$,$n$)$^3$He.  The $^6$Li abundance, however, could show 
a sizable change and therefore the future precise measurement of the cross
sections at high energy $E_\gamma \gtrsim$ 50~MeV is highly required.
\end{abstract}



\maketitle

\section{Introduction}

In standard cosmology, the universe is thought to have experienced big-bang 
nucleosynthesis (BBN) at a very early stage.  The light nuclides D, T+$^3$He, 
$^4$He and $^7$Li+$^7$Be are produced in the standard BBN (SBBN) at
observable levels, while this model does not make appreciable quantities of $^6$Li.  The Wilkinson 
Microwave Anisotropy Probe (WMAP) satellite has measured the temperature 
fluctuations of the cosmic microwave background (CMB) radiation, and 
parameters characterizing the standard big bang cosmology have been 
deduced~\cite{Spergel:2003cb,Spergel:2006hy} from these data. For the 
baryon-to-photon ratio $\eta_{\rm CMB}$ deduced from fits to the CMB, the BBN model 
predicts abundances of the light elements except for $^6$Li  and $^7$Li 
which are more-or-less consistent with those inferred from astronomical 
observations.  

Spectroscopic lithium abundances have been detected in the atmospheres of 
metal poor stars.  Nearly constant abundances of $^6$Li and $^7$Li in metal-poor 
Population II (Pop~II) stars have been inferred. There is about a factor of 
three under-abundance of $^7$Li in metal-poor halo stars (MPHSs) with respect 
to the SBBN prediction when using the baryon-to-photon ratio 
$\eta_{\rm CMB}$.  This is called the $^7$Li 
problem~\cite{Ryan:1999vr,Melendez:2004ni,Asplund:2005yt}.  In addition, spectroscopic measurements obtained with high resolution 
indicate that MPHSs have a very large abundance of $^6$Li, 
i.e.~at a level of about 3 orders of magnitude larger than the SBBN prediction 
of the $^6$Li abundance, which is called the $^6$Li
problem~\cite{Asplund:2005yt,ino05}.  Cayrel et al.~\cite{Cayrel:2007te}
studied line asymmetries to be generated by convective Doppler shifts in
stellar atmospheres, and found that the convective asymmetry might mimic the
presence of $^6$Li and an error of $^6$Li/$^7$Li amounts to a
few percent that is roughly comparable to the values estimated from MPHSs.  The
$^6$Li problem, therefore, may not exist in fact, since the
convective asymmetry could give a possible solution to the $^6$Li problem
within the framework of SBBN.

The possibility of $^6$Li production in non-standard BBN  triggered by the decay 
of unstable relic neutral massive particles $X$ has been studied~\cite{Jedamzik:2006xz,
Jedamzik:1999di,Kawasaki:2000qr,Cyburt:2002uv,Kusakabe:2006hc,Jedamzik:2004er,Jedamzik:2004ip,
Kawasaki:2004qu,Cumberbatch:2007me}. Several critical constraints on the properties of $X$ particles 
were derived from the studies of radiative decay~\cite{Jedamzik:1999di,
Kawasaki:2000qr,Cyburt:2002uv,Kusakabe:2006hc}, hadronic decay or annihilation~\cite{Jedamzik:2004er,
Jedamzik:2004ip,Kawasaki:2004qu,Cumberbatch:2007me} of $X$ particles along with the BBN constraints 
on the light elements. These particle decay induces electromagnetic and/or 
hadronic showers triggering the destruction of preexisting nuclei and the production 
of different nuclear species.  A recent detailed study~\cite{Kusakabe:2006hc} of the radiative decay and its influence on the $^6$Li production has found 
a parameter region of lifetime $\tau_X\sim 10^8-10^{12}$~s and abundance parameter 
$\zeta_X\sim 10^{-13}-10^{-12}$~GeV where the non-thermal nucleosynthesis of $^6$Li 
can explain the observed abundance level in MPHSs. This parameter region satisfies 
the two observational constraints on the CMB energy spectrum and the primordial 
light element abundances. Three important characteristics were found for the 
interesting parameter region. First, $^3$He and $t$ are the seeds for $^6$Li in 
the processes $^4$He($^3$He,$p$)$^6$Li and $^4$He($t$,$n$)$^6$Li. Second, the 
excess of $^6$Li abundance is therefore regulated by the amounts of $^3$He and $t$ 
which are produced by the non-thermal photodisintegration of $^4$He, 
i.e. $^4$He($\gamma$,$p$)$^3$H and $^4$He($\gamma$,$n$)$^3$He. 
Hence, the radiative decay model which results in $^6$Li-production above the 
MPHS abundance level is also reflected by an enhancement of the $^3$He abundance 
with respect to the SBBN value. Third, the radiative decay does not resolve
the $^7$Li problem~\cite{footnote}.  It is therefore concluded that other mechanisms such as the stellar depletion of 
the lithium isotopes in the atmosphere of MPHSs~\cite{Richard:2004pj,Lambert:2004kn} 
or new burst of late-time BBN on the exotic $X$-bound nuclei in the case of 
negatively-charged leptonic particles $X^-$~\cite{Pospelov:2006sc,Hamaguchi:2007mp,Cyburt:2006uv,Kohri:2006cn,Bird:2007ge,Kusakabe:2007fu,Kusakabe:2007fv,Kawasaki:2007xb,Kawasaki:2008qe,Jittoh:2007fr,Jittoh:2008eq,Jedamzik:2007cp,Jedamzik:2007qk,Pospelov:2007js,Pospelov:2008ta} must operate to lower the $^7$Li abundance. 

A recent measurement of $^4$He photodisintegration reactions, 
$^4$He($\gamma$,$p$)$^3$H and $^4$He($\gamma$,$n$)$^3$He with 
laser-Compton photons~\cite{Shima:2005ix} shows much smaller cross sections than 
those estimated from the other previous experiments~\cite{Nakayama2007} and 
those summarized in Ref.~\cite{Cyburt:2002uv} at the photon energies 20~MeV 
$\lesssim E_{\gamma} \lesssim$ 30~MeV. 
If these non-thermal photon energies dominate the destruction of $^4$He, the 
production of $^3$He and $t$ and also the subsequent production of $^6$Li via 
$^4$He($^3$He,$p$)$^6$Li and $^4$He($t$,$n$)$^6$Li, 
this would change the parameter region of $\tau_X$ and $\zeta_X$ of massive 
relic particles $X$ so that the resultant non-thermal nucleosynthesis of $^6$Li 
can explain the abundance level observed in MPHSs. 
The first purpose of this article is to study the sensitivity of non-thermal 
BBN of all light elements D, T, $^3$He, $^4$He, $^6$Li, $^7$Li and $^7$Be to 
the photodisintegration cross sections of $^4$He. 
The second purpose is to infer the uncertainties of the two parameters 
$\tau_X$ and $\zeta_X$ of massive relic particles $X$ which would arise from 
the uncertainties of the measured reaction cross sections. 
 
In Sec.~\ref{sec2} we present a result of a new measurement of $^4$He 
photodisintegration cross sections.  In Sec.~\ref{sec3} we briefly 
explain the model of non-thermal nucleosynthesis and the calculated 
result of the effect of the considered change of the photodisintegration 
cross sections.  In Sec.~\ref{sec4} we summarize our conclusion and 
offer an outlook for measurements of $^4$He photodisintegration. 

\section{$^4$H\lowercase{e}($\gamma$,$p$)$^3$H and 
$^4$H\lowercase{e}($\gamma$,$n$)$^3$H\lowercase{e} Cross 
Sections}\label{sec2}

So far the cross section data of the photodisintegration of $^{4}$He have been 
obtained from the direct photodisintegration experiments as well as the inverse 
radiative capture experiments.  Indirect probes such as the ($p$,$p'$) reaction \cite{Yamagata2006} 
and the ($^{7}$Li,$^{7}$Be) reaction \cite{Nakayama2007} have also been applied 
to investigate the property of the dipole excitations of $^{4}$He. \\

The direct experiments have been performed by detecting either charged fragments 
($p$, $^{3}$H, $^{3}$He) or neutrons from the photodisintegration reactions.
As incident real photon beams, either continuous bremsstrahlung photons or
quasi-monochromatic ones generated with various methods such as the photon
tagging, the positron annihilation in flight, and the laser
Compton-backscattering were used.  Since the energies of the emitted particles 
are small due to high threshold energies of the photodisintegrations of $^{4}$He, 
many efforts have been devoted to detect those particles clearly from the 
backgrounds caused by the incident high-energy $\gamma$-rays. 
In the inverse experiments, the reaction cross sections are much smaller than 
the photodisintegrations because of the difference of the phase space factors, 
and therefore the measurements have been performed with great care to the influence 
of the background $\gamma$-rays as well as the determinations of the experimental 
parameters such as the detector efficiency for high-energy capture $\gamma$-rays, 
the effective target thicknesses, the incident beam intensity, and so on. \\

\begin{figure}
\begin{center}
\includegraphics[width=8.0cm,clip]{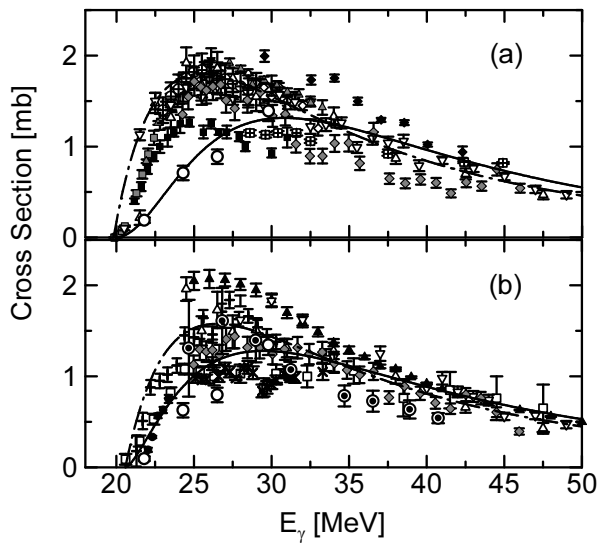}
\caption{The cross sections of the $^{4}$He($\gamma$,$p$)$^{3}$H (upper panel:
  a) and $^{4}$He($\gamma$,$n$)$^{3}$He (lower panel: b) reactions. 
The open circles stand for the published data 
\cite{Shima:2005ix} from experiments performed 
using quasi-monochromatic laser-Compton photon beams at 
AIST.  The dotted circles are the latest 
data of the ($\gamma$,$n$) reaction measured by means of a tagged photon beam 
\cite{Nilsson2007}. 
The other symbols indicate the previous data (see the references in Ref. 
\cite{Shima:2005ix}). The error bars show $1 \sigma$ uncertainties in
 the cross section data.  The solid curves are the most probable 
excitation functions determined from the experimental data at $E_\gamma \lesssim$ 30~MeV~\cite{Shima:2005ix} and the previous ones at
higher energies 30~MeV $\lesssim E_\gamma \lesssim$
116~MeV~\cite{Gorbunov1968,Arkatov1974,Malcom1973}.  The dash-dotted curves are the fitting functions to some of the available data 
summarized in Ref.~\cite{Cyburt:2002uv}.\label{fig1}}
\end{center}
\end{figure}

In spite of the above experimental efforts, there have been large discrepancies 
in the previous data as shown in Fig.~\ref{fig1}. Especially in the energy
region below $\sim$~30~MeV, the data show either a pronounced peak at around 25$-$26 MeV 
or a rather smooth curve as a function of the excitation energy. In order to 
determine the cross sections more accurately, Shima et al. recently performed measurements 
using quasi-monochromatic laser-Compton photons and a time 
projection chamber (TPC) containing a gas mixture of He and CD$_{4}$ 
\cite{Shima:2005ix,Kii2005,shima09}. The method had the following features: \\
$\cdot$ Thanks to the quasi-monochromatic and well-collimated $\gamma$-ray beam, 
the background due to low-energy $\gamma$-rays is very small.\\
$\cdot$ Since the TPC gas serves as an active target, low-energy charged 
fragments from the photodisintegrations can be detected simultaneously for
$^4$He($\gamma$,$p$) and $^4$He($\gamma$,$n$) with efficiencies 
of nearly 100\% and a solid angle of 4$\pi$.\\
$\cdot$ The absolute sensitivity of the measurement can be accurately checked 
with the D($\gamma$,$n$)$p$ reaction, whose cross section is well known in 
the energy region of the present interest. \\

Previously Shima et al. measured the cross sections in the $\gamma$-ray energy region 
from 20 to 30 MeV using the laser-Compton $\gamma$-ray source at the National Institute of 
Advanced Industrial Science and Technology (AIST) \cite{Shima:2005ix}.

As shown in Fig.~\ref{fig1}, the cross sections from the experiments 
are found to monotonically increase as a function of the $\gamma$-ray 
energy up to $\sim$ 30~MeV, being quite different from the standard fitting 
functions previously evaluated by Cyburt et al. \cite{Cyburt:2002uv}. 
With the above mentioned situation in mind, the sensitivity of the 
BBN to the photodisintegration cross sections of 
$^{4}$He was studied in the present paper by using the recent 
experimental results as well as the previously adopted standard fitting 
functions in Ref.~\cite{Cyburt:2002uv}.

The present excitation functions were determined as follows. Since 
no sharp resonance has been observed in the excitation functions of 
the photodisintegration of $^{4}$He in the energy region up to 
$\sim$ 100~MeV, we assumed the excitation functions can be approximated by a fitting function in Ref.~\cite{Cyburt:2002uv}, i.e.,

\begin{equation}
\sigma(E_\gamma)= \sigma_c \frac{\left|Q\right|^a(E_\gamma-\left|Q\right|)^b}{E_\gamma^{a+b}}.
\label{cross21}
\end{equation}
Three parameters, $\sigma_c$, $a$ and $b$ are determined by fitting this
functions to the data measured at AIST~\cite{Shima:2005ix} and the data measured at high energies of $E_\gamma \gtrsim$ 30~MeV~\cite{Gorbunov1968,Arkatov1974,Malcom1973} by means of the 
$\chi^{2}-$minimization method.  This function is suitable for use in numerical calculation because it has no
discontinuity and no divergence at high energies.  In this study we take
published values for errors of cross sections at all fitting procedures.

The most probable fitting functions for the cross sections of the
$^4$He($\gamma$,$p$)$^3$H and $^4$He($\gamma$,$n$)$^3$He reactions turn
  out to be
\begin{equation}
\sigma(E_\gamma)= 128.9~{\rm mb}~\frac{\left|Q\right|^{4.524}(E_\gamma-\left|Q\right|)^{2.512}}{E_\gamma^{4.524+2.512}},
\label{cross11}
\end{equation}
and 
\begin{equation}
\sigma(E_\gamma)= 31.68~{\rm mb}~\frac{\left|Q\right|^{3.663}(E_\gamma-\left|Q\right|)^{1.580}}{E_\gamma^{3.663+1.580}},
\label{cross12}
\end{equation}
respectively, and they are plotted in Fig.~\ref{fig1}.  Reaction $Q$-values
are taken from experiments as $\left|Q\right|=19.8139$~MeV and
$\left|Q\right|=20.5776$~MeV, respectively.  We take these two
cross sections as our recommended cross sections in this paper.

The dash-dotted curves are standard expressions from Ref.~\cite{Cyburt:2002uv}.  Parameter values are given as $\sigma_c=19.5$~mb,
$a=3.5$ and $b=1.0$ for $^4$He($\gamma$,$p$)$^3$H reaction, and
$\sigma_c=17.1$~mb, $a=3.5$ and $b=1.0$ for $^4$He($\gamma$,$n$)$^3$He
reaction.

\begin{figure}
\begin{center}
\includegraphics[width=8.0cm,clip]{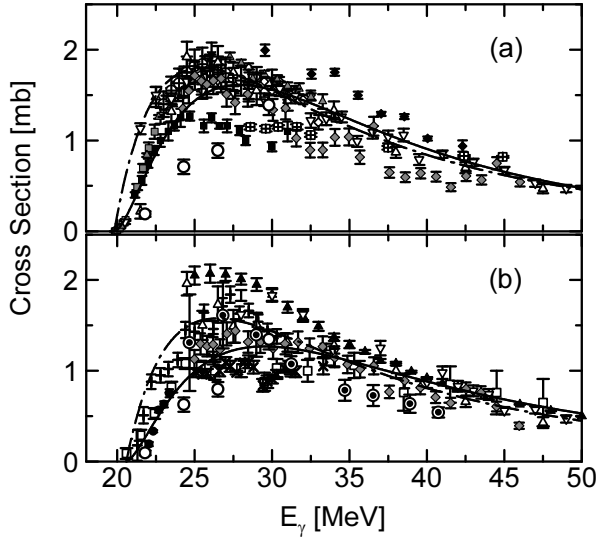}
\caption{(a) and (b) are the same as those in Fig.~\ref{fig1}.  The solid curves and the
short-dashed curves are the most probable excitation functions determined
from all data adopted in this study including the experimental data at
$E_\gamma \lesssim 30$~MeV~\cite{Shima:2005ix} and the previous ones
at whole energy range of $E_\gamma <$
116~MeV~\cite{Nakayama2007,Yamagata2006,Nilsson2007,Gorbunov1968,Arkatov1974,Malcom1973}.
 The dash-dotted curves are the same as in Fig.~\ref{fig1}.\label{fig2}}
\end{center}
\end{figure}

As another implementation, we fit all data adopted in this study
including the data measured at AIST~\cite{Shima:2005ix} at low energies of
$E_\gamma < 30$~MeV and previous ones obtained in other experiment.  The fitting parameters turn out to be $\sigma_c=61.82$~mb,
$a=4.300$, and $b=1.756$ for $^4$He($\gamma$,$p$)$^3$H reaction, and
$\sigma_c=31.38$~mb, $a=3.651$, and $b=1.583$ for $^4$He($\gamma$,$n$)$^3$He
reaction.  The derived cross sections are displayed in Fig.~\ref{fig2}.

One can see in Fig.~\ref{fig1} and Fig.~\ref{fig2} that there is a large
dispersion in experimental data at low photon energies of 20~MeV~$\lesssim
E_\gamma \lesssim 40$~MeV.  This dispersion is much larger than $\pm 1 \sigma$
deviation of any single fitting function.  We therefore adopt the above three
functions in BBN calculations and compare the calculated results with one
another in order to study the sensitivity of BBN with the radiative particle
decay to the $^4$He photodisintegration reaction cross sections.

\section{BBN of the Light Elements}\label{sec3}
\subsection{Model}\label{model}
We assume the creation of high energy photons from the radiative decay of 
a massive particle $X$ with a mass $M_X$ and a mean life of 
$\tau_X=10^{2} - 10^{12}$~s.  We represent the emitted photon energy by 
$E_{\gamma0}$.  We assume that the decaying dark particle is non-relativistic, 
and almost at rest in the expanding universe.

A decay-produced energetic photon interacts with 
the cosmic background and induces an electromagnetic cascade shower. The 
faster processes, pair production through background photons 
$\gamma_{\rm bg}$ ($\gamma\gamma _{\rm bg}\rightarrow e^+ e^-$) and inverse 
Compton scattering of produced electrons and positrons through background photons 
($e^\pm\gamma _{\rm bg}\rightarrow e^\pm \gamma$), produce electromagnetic showers 
and the non-thermal photon spectrum realizes a quasi-static 
equilibrium~\cite{Kawasaki:1994sc,Protheroe:1994dt}. The attained zeroth 
generation photon spectrum is~\cite{berezinskii90},
\begin{equation}
{p_{\gamma}(E_{\gamma})}
\approx \left\{ \begin{array}{ll}
K_0(E_X/E_{\gamma})^{1.5}
&  \mbox{for $E_{\gamma}<E_X$} \\ 
K_0(E_X/E_{\gamma})^{2.0}
&  \mbox{for $E_X\leq E_{\gamma}<E_C$}, \\ 
0
&  \mbox{for $E_C\leq E_{\gamma}$} \\ 
\end{array}
\right.	
\label{spectrum}
\end{equation}
where $K_0=E_{\gamma 0}/\left(E_X^2\left[2+\ln(E_C/E_X)\right]\right)$ is a 
normalization constant fixed by energy conservation of the injected photon 
energy. This spectrum has a break in the power law at $E_\gamma=E_X$ and 
an upper cutoff at $E_\gamma=E_C$. We take the same energy scaling with the 
temperature $T$ of the background photons as in~\cite{Kawasaki:1994sc}, 
i.e.~$E_X=m_e^2/80T$ and $E_C=m_e^2/22T$ at the temperature $T$, where the cascade spectrum was 
calculated by numerically solving a set of Boltzmann equations.  Here $m_e$ is
the rest mass of an electron.

Because the rates of electromagnetic interactions are faster than 
the cosmic expansion rate, the photon spectrum $p_\gamma(E_\gamma)$ is modified into 
a new quasi-static equilibrium (QSE). This distribution is given by 
\begin{equation}
{\mathcal N}_\gamma^{\rm QSE}(E_\gamma) =
\frac{n_Xp_\gamma (E_\gamma)}{\Gamma_\gamma(E_\gamma)\tau_X},
\label{ng}
\end{equation}
where $n_X=n_X^0(1+z)^3 \exp(-t/\tau_X)$ is the number 
density of the decaying particles at a redshift $z$. 
The quantity $\Gamma_\gamma$ is the energy degradation rate through 
three slower processes; Compton scattering 
($\gamma e^\pm_{\rm bg} \rightarrow\gamma e^\pm$), Bethe-Heitler ordinary 
pair creation in nuclei ($\gamma N_{\rm bg}\rightarrow e^+ e^- N$), 
and double photon scattering 
($\gamma \gamma_{\rm bg} \rightarrow \gamma \gamma$) for the 
zeroth-generation photons.  We use this steady state approximation 
for the cosmic non-thermal constituent of photons. 

The equation for the production and destruction of nuclei by 
non-thermal photons is given by 
\begin{equation}
\frac{d Y_A}{d t} = \sum_T N_{AC}\left[T\gamma\right]_A Y_T - \sum_P \left[A\gamma\right]_P Y_A,
\label{dydt2}
\end{equation}
where we have defined the reaction rate 
\begin{eqnarray}
\left[T\gamma\right]_A&
\hspace{-5pt}
=&
\hspace{-5pt}
\frac{n_\gamma^0 \zeta_X}{\tau_X}
 \left(\frac{1}{2H_r t}\right)^{3/2} \exp(-t/\tau_X)
\nonumber \\
&&
\hspace{-5pt}
\times \int_{0}^\infty dE_\gamma
S_\gamma^{\rm QSE}(E_\gamma)\,  \sigma_{\gamma +T \rightarrow A}
(E_\gamma),
\nonumber \\
\label{agamma}
\end{eqnarray}
and
\begin{eqnarray}
S_\gamma^{\rm QSE}(E_\gamma)=\frac{\tau_X}{E_{\gamma 0}n_X}{\mathcal N}_\gamma^{\rm QSE}(E_\gamma).
 \label{sgamma}
\end{eqnarray}
$Y_i\equiv n_i/n_B$ is the mole fraction of a particular nuclear 
species $i$, and $n_i$ and $n_B$ are number densities of nuclei $i$ and 
baryons.  The first and second term on the right-hand side are the 
source ($\gamma+T\rightarrow A+C$) and sink ($\gamma+A\rightarrow P+D$) 
terms for nucleus $A$.  $E_\gamma$ is a non-thermal photon energy. 
The cross section of the process $\gamma + T \rightarrow A+C$ is denoted by 
$\sigma_{\gamma + T \rightarrow A}(E_\gamma)$. Further we use $N_{AC}$ to 
represent the number of identical species of nuclei in a production or 
destruction process; $N_{AC}=2$ when particles $A$ and $C$ are identical 
and $N_{AC}=1$ when they are not.  For example, in the process 
$^4$He($\gamma$,$d$)D, $N_{\rm DD}=2$. 
We defined $H_r\equiv \sqrt[]{\mathstrut 8\pi G \rho_{\rm rad}^0/3}$, 
where the superscript 0 denotes present values ($z=0$), therefore 
$n_\gamma^0$ and $\rho_{\rm rad}^0$ are the present photon number density 
and present radiation energy density of the cosmic background radiation 
(CBR), respectively. We defined $\zeta_X=(n_X^0/n_\gamma^0)E_{\gamma0}$.

The equation describing the secondary production and destruction is 
obtained by taking account of the energy loss of nuclear species while 
propagating through the background. In general, because of the high energy 
loss rate, the primary particles establish a quasi-static equilibrium. 
The abundance evolution is then represented by 
\begin{equation}
\frac{d Y_S}{d t} = \sum_{T,A,T'}Y_T Y_{T'} \frac{N_{AX_1} N_{SX_2}}{N_{AT'}}
 [T(A)T']_S -({\rm sink~term}),
\label{dysdt}
\end{equation}
where $[T(A)T']_S$ is the reaction rate for a secondary process 
$T(\gamma,X_1)A(T',X_2)S$ with any combination of particles 
$X_1$, $A$, and $X_2$. 
For example, $[\alpha(^3$He$) \alpha$]$_{^6{\rm Li}}$ for a secondary process 
$^4$He($\gamma$,n)$^3$He($\alpha$,$p$)$^6$Li is given by 
\begin{eqnarray}
\left[\alpha \alpha \right]_{^6{\rm Li}} 
\hspace{-5pt}
&=&
\hspace{-5pt}
 \frac{\eta (n_\gamma^0)^2 \zeta_X}{\tau_X}
 \left(\frac{1}{2H_r t}\right)^3 \exp(-t/\tau_X)
\nonumber \\
&&
\hspace{-5pt}
\times \int_{E_{p,\rm th}}^{{\mathcal E}_{^3{\rm He}}(E_C)}\!\!\!\!\! dE_{^3{\rm He}} 
 \frac{\sigma_{^3{\rm He}(\alpha,p)^6{\rm Li}}(E_{^3{\rm He}})\beta_{^3{\rm He}}}{b_{^3{\rm He}}(E_{^3{\rm He}})}
\nonumber \\
&&
\hspace{-5pt}
\times \int_{{\mathcal E}_{^3{\rm He}}^{-1}(E_{^3{\rm He}})}^{E_C} \!\!\!\!\! dE_\gamma
 S_\gamma^{\rm QSE}(E_\gamma)\sigma_{^3{\rm He}(\alpha,p)^6{\rm Li}}
 (E_\gamma),
\nonumber \\
 \label{tt'}
\end{eqnarray}
where $\eta$ is the baryon-to photon ratio:~$\eta\equiv n_B^0/n_\gamma^0$, 
and $\beta_A$ is the velocity of the primary particle $A$. $b_A=-dE/dt$ is 
the energy loss rate of the primary particle by Coulomb scattering. 
${\mathcal E}_A(E_\gamma)$ is the energy of the nuclide $A$ produced by 
the photodisintegration process $\gamma+T\rightarrow A$, and 
${\mathcal E}_A^{-1}(E_A)$ is the energy of the non-thermal photons which 
produce the primary species $A$ with energy $E_A$.  ${\mathcal E}_A(E_\gamma)$ 
and ${\mathcal E}_A^{-1}(E_A)$ are derivable in the limit of low energy 
scattering, where the relevant nuclei are non-relativistic, i.e. 
${\mathcal E}_{^3{\rm He}}(E_\gamma)=(E_\gamma-E_{\gamma,{\rm th}})/4$ and 
${\mathcal E}_{^3{\rm He}}^{-1}(E_{^3{\rm He}})=4E_{^3{\rm He}}+E_{\gamma,{\rm th}}$, 
where $E_{\gamma,{\rm th}}$ is the threshold energy of primary photodisintegration 
reaction.

\subsection{Constraints on the radiative decay of long-lived
  particles}\label{result}

\begin{figure}[tbp]
\includegraphics[width=8.0cm,clip]{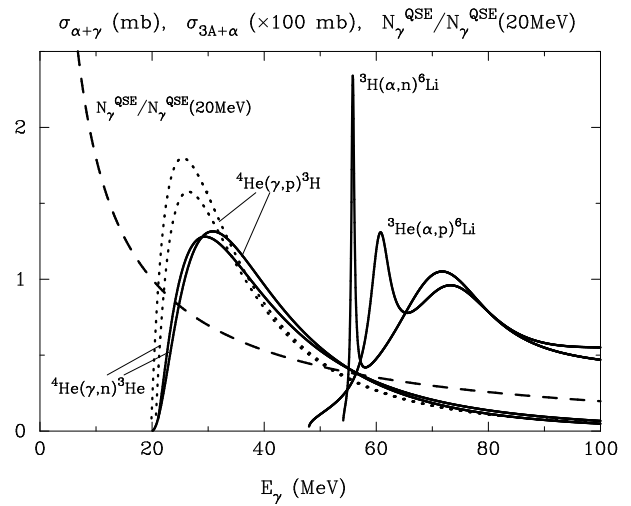}
\caption{The fitted cross sections of $^4$He($\gamma$,$n$)$^3$He and 
$^4$He($\gamma$,$p$)$^3$H with data from the laser-Compton photon 
experiment at low energies of $E_\gamma \lesssim 30$~MeV are shown by solid
lines above $\sim 20$~MeV. The standard cross sections~\cite{Cyburt:2002uv} are also shown as dotted lines. Cross 
sections of secondary $^6$Li production reactions $^3$H($\alpha$,$n$)$^6$Li 
and $^3$He($\alpha$,$p$)$^6$Li are superimposed. The energy spectrum of 
non-thermal photon produced by the radiative decay when the cosmic 
temperature is 10~eV corresponding to the decay life $\tau_X \sim 10^{10}$~s 
is also shown (dashed line), which is normalized with respect to intensity 
at $E_\gamma=20$~MeV.\label{fig5}}
\end{figure}

We focus on the non-thermal production of mass 3 nuclides, $^3$H and 
$^3$He, and secondary production of $^6$Li. In Fig.~\ref{fig5}, 
photodisintegration cross sections of $^4$He($\gamma$,$n$)$^3$He and 
$^4$He($\gamma$,$p$)$^3$H are shown as a function of non-thermal 
photon energy $E_\gamma$ (above threshold energy $20$~MeV).  These 
cross sections are fitted with data from the laser-Compton photon 
experiment (solid curves) and those from Ref.~\cite{Cyburt:2002uv} 
(dotted curves).  Dashed line is the energy spectrum 
of non-thermal photon produced by the radiative decay when the cosmic 
temperature is 10~eV corresponding to the decay life $\tau_X \sim 
10^{10}$~s~\cite{Kusakabe:2006hc}.  It is normalized with respect to intensity at 
$E_\gamma=20$~MeV. Since the yield of non-thermally produced mass 3 
nuclides is proportional to the integration of the non-thermal photon 
spectrum times photodisintegration cross sections, the change of 
photodisintegration cross sections at $E_\gamma \lesssim 30$~MeV has 
relatively large influence on the resulting $^3$He abundance. 


On the other hand, non-thermal secondary production of $^6$Li is not
  affected by $^4$He photodisintegration cross sections at low energies. 
When we consider the non-relativistic limit of nuclear reactions, the center of 
mass energy $E_{\rm cm}$ in $^3$He($\alpha$,$p$)$^6$Li and $^3$H($\alpha$,$n$)$^6$Li 
reactions is given by 
\begin{equation}
 E_{\rm cm}=\frac{1}{2}\mu v^2 \sim \frac{1}{2} \frac{m_3 m_4}{m_3+m_4} v_3^2
 \sim \frac{4}{7}E_3,
  \label{eq1}
\end{equation}
where $\mu$, $m_3$ and $m_4$ are the reduced mass, the mass of a nucleus $^3A$
($t$ or $^3$He) and that of $^4$He, $v$ and $v_3$ are the relative velocity
and the velocity of $^3A$, $E_3$ is the kinetic energy of $^3A$,
respectively.  The energy of $^3A$ generated in primary $^4$He($\gamma$,$n$)$^3$He and
$^4$He($\gamma$,$p$)$^3$H reactions, i.e., $E_3^0$ ($\geq E_3$) is
given by
\begin{equation}
 E_3^0\sim \frac{m_p}{m_3+m_p} (E_\gamma-E_{\gamma,{\rm th}}) \sim \frac{1}{4} (E_\gamma-E_{\gamma,{\rm th}}),
  \label{eq2}
\end{equation}
where $m_p$ is the mass of proton and $E_{\gamma,{\rm th}}$ is the threshold energy of $^4$He 
photodisintegration reaction.  The center of mass energy $E_{\rm cm}$ in the
secondary reactions thus relates to the non-thermal photon energy in primary reactions as 
\begin{equation}
 E_{\rm cm}\lesssim \frac{E_\gamma-E_{\gamma,{\rm th}}}{7}.
  \label{eq3}
\end{equation}
The inequality means that non-thermal 
$^3$He and $^3$H nuclides lose energies from a time of their production 
to a time of nuclear reactions to produce $^6$Li. We plot the cross sections 
of $^3$He($\alpha$,$p$)$^6$Li and $^3$H($\alpha$,$n$)$^6$Li reactions as a 
function of $E_\gamma$ neglecting the energy losses of $^3$He and $^3$H in Fig.~\ref{fig5}.  Since secondary $^6$Li production needs photon energy
of at least $E_\gamma \gtrsim 50$~MeV to produce energetic $^3$He and $^3$H, the yield of $^6$Li is 
insensitive to the change of the $^4$He photodisintegration cross section at 
low energy of $E_\gamma \lesssim 30$~MeV.  (See Sec.~\ref{sec3}B1.)  On the other hand, a
change of the cross section at high energies of $E_\gamma \gtrsim$~50~MeV causes a change of resulting yield of $^6$Li  (See Sec.~\ref{sec3}B2 for estimation of uncertainty on the $^6$Li yield associated with a possible
uncertainties in cross sections at high energies.)

\subsubsection{Effect of uncertainties from low energy reactions}

\begin{figure}[tbp]
\includegraphics[width=8.0cm,clip]{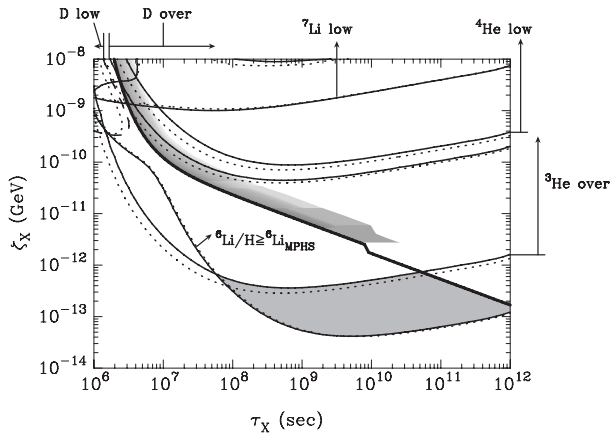}
\caption{Contours in the $(\tau_X,\zeta_X)$ plane 
corresponding to the adopted constraints for the primordial abundances 
in the present calculation with smaller cross sections of $^4$He 
photodisintegration Eqs.~(\ref{cross11}) and (\ref{cross12}) (thick lines). Contours for the mass fraction of $^4$He 
$Y=0.232$ and the number ratios of $^3$He/H=$3.1\times 10^{-5}$, 
D/H=$5.2\times 10^{-5}$, D/H=$1.4\times 10^{-5}$ (dashed line), and 
$^7$Li/H=$1.1\times 10^{-10}$ are shown. The notation ``over'' and 
``low'' identifies overproduced and underproduced regions, 
respectively. The same constraints on the result calculated with 
larger standard cross sections of $^4$He photodisintegration~\cite{Cyburt:2002uv} 
is shown as dotted lines. The region above the thick solid line is 
excluded by the consistency requirement of the CMB with a blackbody. 
The contour of $^6$Li/H=$6.6\times 10^{-12}$ is also drawn. 
The gray region above the contour and below the nucleosynthesis plus 
CMB constraints is the allowed region where the abundant $^6$Li is produced.\label{fig3}}
\end{figure}

Figure~\ref{fig3} shows contours corresponding to the constraints for the 
primordial abundance~\cite{Kusakabe:2006hc} for the present result with our
recommended smaller cross sections of $^4$He photodisintegration Eqs.~(\ref{cross11}) and (\ref{cross12}) (thick lines).  For example, a contour of the $^4$He mass fraction $Y>0.232$ is shown in the 
($\tau_X$,$\zeta_X$) plane. The region above this contour should be 
excluded for $Y<0.232$. The contours of $^7$Li/H lower limit, 
$^7$Li/H$>1.1\times10^{-10}$, D/H upper and lower limits, 
D/H$ \leq5.2\times 10^{-5}$ and D/H$ \geq1.4\times 10^{-5}$ (dashed line), 
respectively, and $^3$He/H upper limit, $^3$He/H$ \leq 3.1\times 10^{-5}$ 
are also drawn. Dotted lines show contours of the same constraints for the 
result with larger standard cross sections of $^4$He photodisintegration~\cite{Cyburt:2002uv}. 
The thick solid line shows a constraint from the consistency requirement of 
the CMB with a blackbody~\cite{Kusakabe:2006hc}. The region above the 
line is excluded. The contour for the MPHSs level of $^6$Li/H=$6.6\times 10^{-12}$ 
is plotted. The gray region above the contour and below the nucleosynthesis 
plus CMB constraints is allowed and abundant in $^6$Li.


For this figure, we use the two sigma upper limit on the observed $^3$He/H
abundance ratio from Galactic HII region~\cite{ban02} as a conservative constraint on
primordial abundance.  Alternatively a constraint from $^3$He/D ratio is
often used.  Since deuterium is fragile and more easily burned in stars than $^3$He is,
the $^3$He/D ratio is thought to increase monotonically as a function of time by stellar
processing from the formation of first stars to the solar system formation.
The $^3$He/D ratio for primordial abundances can, therefore, be constrained
by the solar $^3$He/D ratio.  Since the solar $^3$He/D ratio is ($^3$He/D)$_\odot$=0.82~\cite{lod03}, one can obtain a constraint of $^3$He/D $<1$ for
primordial abundances.  Although this constraint is slightly stronger than the
constraint for the $^3$He/H ratio, the contours for both constraints
are not distinguishable from each other in Fig.~\ref{fig3}.

One can see that $^7$Li abundance does not change drastically by the new 
$^4$He photodisintegration cross sections. The contour for the $^4$He 
over-destruction by the non-thermal photons shifts upward by $\sim$ 
300~\% at most for $\tau_X=10^6-10^{10}$~s with respect to that 
with the larger cross sections (dotted line). Since a destruction rate 
of $^4$He is lowered at low energies when its cross sections are taken 
to be small, larger energy is necessary to destruct $^4$He considerably. 
The $^4$He destruction leads to non-thermal production of $^3$He and D. 
Then, D over-destruction regions, which are above the region of this figure, 
shift upward reflecting the change of relic abundance of seed nuclide $^4$He. 
The $^3$He overproduction region also shifts upward by $\sim$ 30~\%. 
Since the cross sections of $^4$He($\gamma$,$p$)$^3$H and 
$^4$He($\gamma$,$n$)$^3$He are lowered, the amount of produced mass 3 
nuclides are lowered. One finds an upward shift of D overproduction 
region. We confirmed that this change is caused since D is produced by 
$p$($n$,$\gamma$)D using non-thermally produced neutrons by 
$^4$He($\gamma$,$n$)$^3$He, whose cross section changed. 
Figure~\ref{fig3} shows that the contour of $^6$Li abundance also 
changes slightly by $\sim -20$~\% at most for $\tau_X=10^6 -10^{10}$~s, but 
this change does not result from the change of photodisintegration cross 
sections of $^4$He at low energy $E_\gamma \lesssim$~30~MeV. It is due 
to the difference of excitation functions between our recommended fit and 
the standard fit~\cite{Cyburt:2002uv} at higher energy, $E_\gamma \gtrsim$ 30~MeV.  (See Fig.~\ref{fig5}.)  The reason has already been explained at the beginning
of this section Sec.~\ref{sec3}B.

\begin{figure}[tbp]
\includegraphics[width=8.0cm,clip]{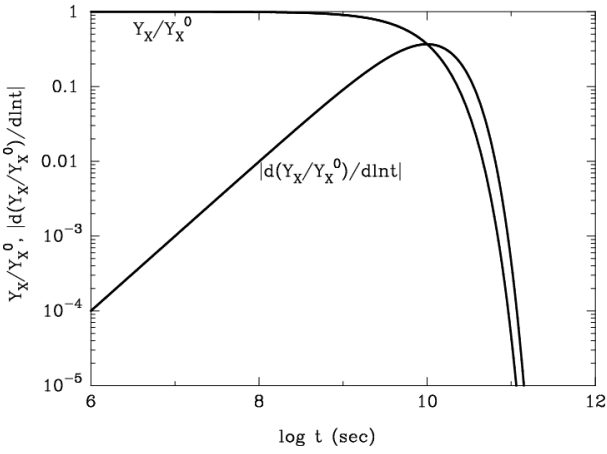}
\caption{Time evolution of abundance of $X$ particles 
$Y_X$/$Y_X^0$, where $Y_X^0$ is the initial abundance before 
$X$-decays. Also shown is 
$t\times \left|d(Y_X/Y_X^0)/dt\right|=(t/\tau_X)\exp(-t/\tau_X)$, 
which characterizes the amount of energy from the particle decay 
at time $t$. The decay life is assumed to be $\tau_X=10^{10}$~s.\label{fig4}}
\end{figure}

We pick up the most interesting parameter region, where $^6$Li is 
produced at the level higher than that of MPHSs. We take a look at the 
result of nucleosynthesis of the parameters $(\tau_X,\zeta_X)$=(10$^{10}$~s, 
$3\times 10^{-13}$~GeV)~\cite{Kusakabe:2006hc} for this reason. 
In Fig.~\ref{fig4}, the abundance of $X$ particle $Y_X$ is shown as a 
function of the cosmic photon temperature. It is normalized as $Y_X$/$Y_X^0$, 
where $Y_X^0$ is the initial abundance before the $X$-decay decreases its 
abundance. As the indicator of energy amount injected by the $X$ decay at 
an epoch, we also plot $\left|d(Y_X/Y_X^0)/d\ln t\right|=t\times
\left|d(Y_X/Y_X^0)/dt\right|=(t/\tau_X)\exp(-t/\tau_X)$. Most of the energy 
content is injected in the time scale of $\tau_X$. These lines are for only 
the case of $\tau_X=10^{10}$~s.


\begin{figure}[tbp]
\includegraphics[width=8.0cm,clip]{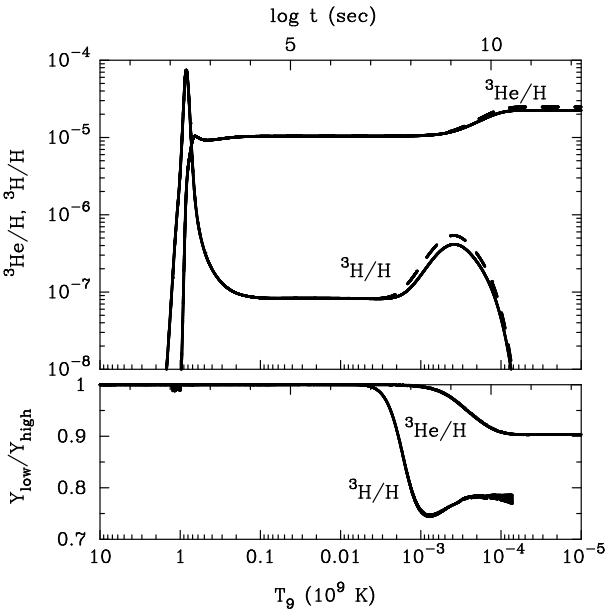}
\caption{(upper panel) Temperature (time) evolution of $^3$He and $^3$H 
abundances when parameters are $(\tau_X, \zeta_X)$=(10$^{10}$~s, 
$3\times 10^{-13}$~GeV).  The solid and dashed lines are the results for our
recommended and the standard cross sections of $^4$He photodisintegration.  (lower panel) Ratios of $^3$He and $^3$H 
abundances calculated with our recommended cross sections to those with the
standard cross sections.\label{fig6}}
\end{figure}

Figure~\ref{fig6} shows the $^3$He and $^3$H abundances as a function of 
temperature $T$ [$T_9\equiv T/(10^9~{\rm K})$] when parameters are 
$(\tau_X, \zeta_X)$=(10$^{10}$~s, $3\times 10^{-13}$~GeV) (upper panel). 
The results with our recommended and the standard cross sections of $^4$He
photodisintegration correspond to the solid and dashed lines, respectively.
An apparent reduction of non-thermally produced $^3$He abundance is found in our recommended case.  The resulting difference of non-thermal 
yield of $^3$He is about 10~\%.


\subsubsection{Effect of uncertainties from high energy reactions}

The measured $^4$He photodisintegration cross sections are still
subject to large uncertainties which depend on the different
experimental setup and methods over the wide energy range. The energy
dependence of the cross sections is not well understood
theoretically. If either cross section of $^4$He($\gamma$,$p$)$^3$H or
$^4$He($\gamma$,$n$)$^3$He in the energy region of $E_\gamma \gtrsim 50$~MeV is different from the adopted
cross section, non-thermal BBN abundance of $^6$Li would change
significantly.  

It should be noted here that the data below 30 MeV of Shima et al.~\cite{Shima:2005ix} are
significantly smaller than previous ones, and therefore one may expect
enhancement of the cross sections in the higher energy range from the
well-known Thomas-Reiche-Kuhn (TRK) sum rule, which relates the
energy-integrated cross section $\sigma_0$ to the ground state property of
a target nucleus.  The E1 sum rule is
expressed by
\begin{eqnarray}
 \sigma_0&=&\int_{E_{\rm th}}^\infty \sigma(E_\gamma) dE_\gamma =\sigma_{\rm
   TRK} (1+\kappa) \nonumber \\
&=& 59.74(1+\kappa)~{\rm MeV mb},
  \label{e1sum}
\end{eqnarray}
where $\sigma_{\rm TRK}=(2\pi^2 \alpha/m)(NZ/A)$ with $\alpha$, $m$, $N$, $Z$
and $A$ the fine structure constant, the nucleon mass, the neutron number, the
proton number and the nucleon number, respectively.  $\kappa$ is the so-called TRK
enhancement factor defined as
\begin{equation}
 \kappa=\frac{mA}{NZ} \langle 0|[D_z, [V, D_z] ] |0\rangle,
  \label{kappa}
\end{equation}
where $|0\rangle$ is the nuclear ground state wave function, $D_z$ is the dipole
operator $D_z=\sum_{i=1}^A z_i
\tau_i^3/2$ with $\tau_i^3$ and $z_i$ the third component of the isospin
operator and the $z$-coordinate of the $i$th particle in the center-of-mass
frame, respectively, $V$ is the nuclear potential.  There is another sum rule,
i.e. bremsstrahlung sum rule, which is expressed by
\begin{eqnarray}
 \sigma_B&=&\int_{E_{\rm th}}^\infty \frac{\sigma(E_\gamma)}{E_\gamma} dE_\gamma
 =4\pi^2 \alpha \langle 0|D_z D_z|0\rangle \nonumber\\
 &=& \frac{4\pi^2 \alpha}{3} \frac{NZ}{A-1} \left(\langle r_\alpha^2\rangle
 -\langle r_p^2\rangle \right)
 \nonumber \\
 &=& 2.60\pm 0.01~{\rm mb},
  \label{bremsum}
\end{eqnarray}
where $\langle r_p^2\rangle^{1/2}=0.875\pm 0.007$~fm~\cite{yao06} and $\langle r_\alpha^2\rangle^{1/2}=1.673\pm
0.001$~fm~\cite{borie78} are the root mean square charge radii for proton and
$^4$He, respectively.

Energy-weighted integrals $\int_{E_{\rm th}}^\infty (E_\gamma)^n
\sigma(E_\gamma) dE_\gamma$ with $n=0$ for Eq.~(\ref{e1sum}) and $n=-1$ for
Eq.~(\ref{bremsum}) are listed in Table~\ref{tab1} for the four different
models of photodisintegration cross sections of $^4$He($\gamma$,$p$)$^3$H and
$^4$He($\gamma$,$n$)$^3$He.  The three models of our recommended fit, i.e.,
Eqs.~(\ref{cross11}) and (\ref{cross12}), the fit with all data adopted in
this study, and the standard fit~\cite{Cyburt:2002uv} lead to values smaller than the sum rules,
  i.e. $\sigma_0=59.7(1+\kappa)$~MeV mb with $\kappa\approx 1$ for the TRK sum rule and $\sigma_B=2.60\pm
0.01$~mb for the bremsstrahlung sum rule.  We expect that the model using
Eqs.~(\ref{cross11}) and (\ref{cross12}) is closer to the lower limit of true
cross sections in consideration of two sum rules.

\begin{table*}
\caption{\label{tab1} Energy-weighted Integrals of Photodisintegration Cross
  Sections $\sigma(E_\gamma)$ in Eqs. (\ref{e1sum}) and (\ref{bremsum})}
\begin{ruledtabular}
\begin{tabular}{c|c|c|c|c}
Models for $\sigma(E_\gamma)$ & Eqs.~(\ref{cross11}), (\ref{cross12}) &
Fit with All Data\footnotemark[1] & Cyburt et
al.~\cite{Cyburt:2002uv}\footnotemark[2] &  Sum Rule\footnotemark[3] \\\hline
$\sigma_0$ [MeV mb] & 77.7 & 79.7 & 84.4 & 118.\footnotemark[4]\\\hline
$\sigma_B$ [mb] & 1.94 & 2.08 & 2.32 & 2.60\\
\end{tabular}
\footnotetext[1]{Eq.~(\ref{cross21}) with
  $\left|Q\right|=19.8139$~MeV,$\sigma_c=61.82$~mb, $a=4.300$, and $b=1.756$
  for $^4$He($\gamma$,$p$)$^3$H, and $\left|Q\right|=20.5776$~MeV,
  $\sigma_c=31.38$~mb, $a=3.651$, and $b=1.583$ for
  $^4$He($\gamma$,$n$)$^3$He.}
\footnotetext[2]{Eq.~(\ref{cross21}) with $\left|Q\right|=19.8139$~MeV,
  $\sigma_c=19.5$~mb, $a=3.5$, and $b=1.0$ for $^4$He($\gamma$,$p$)$^3$H, and
  $\left|Q\right|=20.5776$~MeV, $\sigma_c=17.1$~mb, $a=3.5$, and $b=1.0$ for
  $^4$He($\gamma$,$n$)$^3$He.}
\footnotetext[3]{The functions derived by fitting Eq.~(\ref{cross21}) to
  the data obtained at AIST~\cite{Shima:2005ix} in $E_\gamma \lesssim$
  29.8~MeV are modified at higher energies so that the energy-weighted
  integrals satisfy two sum rules;  Eq.~(\ref{cross21}) with
  $\sigma_c=21.0$~mb, $a=1.68$, and $b=1.88$ for $^4$He($\gamma$,$p$)$^3$H,
  and $\sigma_c=16.8$~mb, $a=1.41$, and $b=1.73$ for
  $^4$He($\gamma$,$n$)$^3$He (for $E_\gamma \leq$ 29.8~MeV) and
  $\sigma(E_\gamma)=2.65~{\rm mb}~(E_\gamma/29.8~{\rm MeV})^{-5/2}$ (for
  $E_\gamma >$ 29.8~MeV).}
\footnotetext[4]{Value derived from $\sigma_0=59.7(1+\kappa)$ with
  $\kappa\approx 1$~\cite{okamoto62,arima73,Gazit:2006ey,weng73}.}
\end{ruledtabular}
\end{table*}

In addition to the three models in the second, third and forth columns in
Table~\ref{tab1} above mentioned, we make another new fitting function of the
cross sections so that the energy weighted integrals satisfy the two sum
rules:  As for low energy cross sections at $E_\gamma \lesssim$
30~MeV, we fit the experimental data~\cite{Shima:2005ix} measured by using
quasi-monochromatic laser-Compton photon beams at AIST.  At higher energy $E_\gamma \gtrsim$~30~MeV, we assume a simple energy
dependence $\sigma(E_\gamma)=\sigma_C(E_\gamma/29.8~{\rm MeV})^{-N/2}$.
Resultant parameters are $\sigma_c=2.65$~mb and $N=5$.  In this new model where we require the sum rules, the constructed cross sections are
typically $\gtrsim 2$ times larger than our recommended ones at 50~MeV $\lesssim
E_\gamma \lesssim$ 135~MeV, where the upper limit is the meson mass.  The both
TRK sum rule and bremsstrahlung sum rule do not apply to such high energies $E_\gamma \gtrsim$~135~MeV because new degrees of freedom of mesons as well as
nucleons play an important role and the above two sum rules break down.  We can expect
that the constructed cross sections which satisfy the sum rules are close to
the upper limits to the realistic cross sections.

\begin{figure}[tbp]
\includegraphics[width=8.0cm,clip]{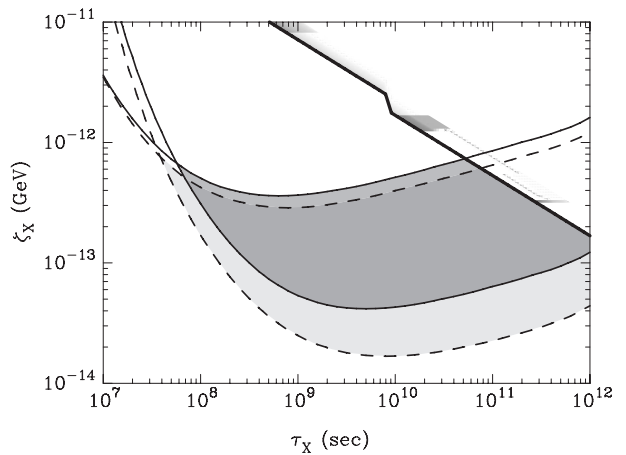}
\caption{Contours in the $(\tau_X,\zeta_X)$ plane 
corresponding to the adopted constraints for the primordial abundances 
in the calculation with our recommended cross sections of $^4$He 
photodisintegration (thick lines) and those satisfying sum rules (dashed
lines).  The adopted abundance constraints on light elements are the same as those in Fig~\ref{fig3}.
 The dark (for our recommended cross sections) and light (for larger ones) gray
 regions above the solid and dashed contour lines, respectively, and below the nucleosynthesis plus CMB
 constraints are the allowed region where abundant $^6$Li is produced.\label{fig7}}
\end{figure}

We compare the primordial abundance calculated with new cross sections
  which satisfy the sum rules and that of our recommended cross sections.  In
Fig.~\ref{fig7} thick solid and dashed lines correspond to the results with
our recommended and new cross sections of $^4$He photodisintegration,
respectively.  The dark and light gray regions above the solid and dashed contour lines and below the nucleosynthesis plus CMB
constraints are the allowed region where the abundant $^6$Li is produced.  The
$^6$Li production occurs more efficiently in the case with larger cross sections than with smaller
ones.  This is because the $^4$He
destruction is more effective for larger cross sections at higher energies so
that the more abundant energetic mass 3 nuclides which synthesize $^6$Li are
produced.  For this reason the $^6$Li abundance produced by using the large
$^4$He-photodisintegration cross sections which satisfy the sum rules are
presumed to be a maximum yield from the viewpoint of nuclear structure
physics.  The allowed region of the properties of relic $X$ particle in the
$(\tau_X, \zeta_X)$ plane, therefore, should not move even below the light
gray region in Fig~\ref{fig7}, and we expect that the reality is located
between the dark and light gray regions.

However, it has been known that $\sigma_0$ highly
depends on various effects of residual interactions among nucleons such as the meson-exchange currents~\cite{gari74}, the tensor correlations~\cite{okamoto62,arima73}, the short-range interactions~\cite{Gazit:2006ey},
and so on. In the case of $^4$He, the calculated values of $\sigma_0$
, which were integrated out up to $E_\gamma \sim 135$~MeV, have uncertainty of more than
$\pm 10$~\%,
depending on the nucleon-nucleon potentials and the nuclear models for
the ground state of $^4$He~\cite{Gazit:2006ey,weng73,fink74,elmi83}. In summary, both the existing
experimental data and the theoretical calculations for the $^4$He photodisintegration
cross section in the energy range up to $\sim 135$~MeV contain large
uncertainties, and therefore a precise measurement of the $^4$He
photodisintegration cross section at $E_\gamma \gtrsim 30$~MeV is highly
desirable.  Comprehensive theoretical study of the nuclear structure and
reactions of $^4$He are also necessary in order to clarify many unresolved
nuclear effects and also refine the applicability of an empirical formula such
as Eq.(\ref{cross21}).  These nuclear physics studies would be important to constrain the
life time and the abundance of long-lived relic $X$ particle more precisely.

\section{Summary and Outlook}\label{sec4}

A recent measurement of $^4$He photodisintegration reactions, 
$^4$He($\gamma$,$p$)$^3$H and $^4$He($\gamma$,$n$)$^3$He with 
laser-Compton photons shows lower cross sections at low energies than 
those estimated by other previous experiments. We studied the sensitivity 
of non-thermal BBN of all light elements D, T, $^3$He, $^4$He, $^6$Li, 
$^7$Li and $^7$Be to the photodisintegration cross section of $^4$He. 

The change of cross sections of $^4$He photodisintegration has an 
influence on the non-thermal yields of light elements, D, $^3$He and 
$^4$He, which are related to the photodisintegration cross sections at low 
energy ($\sim 30$~MeV). The upper limit of allowed regions of $X$-abundance 
parameter $\zeta_X$ for these light nuclei shifts upward by $\sim 300-
  30$~\% for $\tau_X=10^6 -10^{10}$~s for this change of the cross sections. 
This arises from the upshift of $^3$He abundance contour (See Fig.~\ref{fig3}).  
On the other hand, the non-thermal $^6$Li production is not very sensitive to 
the change of cross sections at low energy, since the non-thermal secondary 
synthesis of $^6$Li needs energetic photons of $E_\gamma \gtrsim 50$~MeV.

The non-thermal nucleosynthesis triggered by the radiative particle 
decay is one of candidates of the production mechanism of $^6$Li 
observed in MPHSs. In the interesting parameter region of 
$10^8$~s$\lesssim\tau_X \lesssim 10^{12}$~s and $5\times 10^{-14}$~GeV 
$\lesssim \zeta_X \lesssim 5\times 10^{-13}$~GeV which satisfies the $^6$Li 
production above the abundance level observed in MPHSs, 
the lowering of the photodisintegration cross sections at low energy 
$E_\gamma \lesssim 30$~MeV as measured in the recent experiment using 
laser-Compton photons leads to $\sim10$~\% reduction of resulting $^3$He 
abundance, whereas the $^6$Li abundance does not change for the change of 
the cross sections of $^4$He($\gamma$,$p$)$^3$H and $^4$He($\gamma$,$n$)$^3$He. 

Let us briefly discuss other impacts of such a precise cross section measurement. 
Clarifying the effects of photodisintegrations of $^4$He will affect more strongly 
the $\nu$-process in core-collapse supernova (SN) explosions through the 
neutrino-nucleus interactions specifically of $\nu$+$^4$He. 
The weak transition rates for $^4$He($\nu,\nu$'), 
$^4$He($\nu_e$,$e^-$), and $^4$He($\bar{\nu}_e$,$e^+$) are determined similarly 
to the giant electric dipole resonance observed in the photodisintegrations 
with the help of theoretical calculation~\cite{Suzuki:2006qd}. In fact, several 
experiments of measuring the $^4$He photodisintegration cross 
sections~\cite{Shima:2005ix} were carried out for this 
purpose. The precise knowledge of the $^4$He($\nu$,$\nu$'$p$), ($\nu$,$\nu$'$n$), 
($\nu_e$,$e^- p$), and ($\bar{\nu}_e$,$e^+ n$) cross sections is required to 
determine the unknown parameters for neutrino oscillations through the MSW effect 
on the $^7$Li and $^{11}$B production triggered by the $\nu$+$^4$He 
reactions~\cite{Yoshida:2006qz,Yoshida:2006sk}. The energy range 
$E_\nu = 10 - 25$ MeV is very important for the $\nu$-process nucleosynthesis 
in SNe. The mean neutrino energy of SN neutrinos is presumed to be about 
10 - 25 MeV in numerical simulations of the neutrino transfer in core-collapse 
SNe, and the threshold energies for all neutrino-induced spallation reactions 
of $^4$He are $\sim$20 MeV. Therefore, the difference between the newly 
measured~\cite{Shima:2005ix} and previous $^4$He photodisintegration cross sections at 20~MeV 
$\lesssim E_{\gamma} \lesssim$ 30~MeV could be critical. As a result, the 
absolute yields of $^7$Li and $^{11}$B produced in the $\nu$-process in 
core-collapse SNe would be different from one another, depending on the 
assumed $\nu$-process reaction rates as demonstrated 
theoretically~\cite{Suzuki:2006qd,yoshida08} although the ratio of $^7$Li/$^{11}$B 
does not change largely. 

Another recent focus of photodisintegration of $^4$He is on the mechanism of the core-collapse SNe. 
Most SN simulations still do not succeed in the SN explosion in spite of detailed 
numerical studies of the neutrino transfer calculations inside the core. 
Haxton~\cite{Haxton:1988in} proposed that the neutrino-induced excitations of 
$^4$He and heavier nuclei could deposit extra-energy to the ejected materials and 
revive the shock wave, which motivated a recent theoretical study on the role of 
$^4$He spallation reactions in the core-collapse SNe~\cite{Ohnishi:2005cv}. 
His theoretical suggestion also motivated recent experimental studies of 
photodisintegrations of $^4$He~\cite{Shima:2005ix} in order to 
estimate the neutrino-induced reaction cross sections for $^4$He($\nu,\nu$'), $^4$He($\nu_e$,$e^-$), and $^4$He($\bar{\nu}_e$,$e^+$).

As such, it is important and even critical to study the $^4$He($\gamma$,$p$) and 
$^4$He($\gamma$,$n$) reactions precisely for the discussions of the problem of 
SN-neutrino oscillation and SN explosion as well as the cosmological discussion 
concerning the BBN with a radiative decay of long-lived relic particles.

\begin{acknowledgments}
This work has been supported in part by the Mitsubishi Foundation, the 
Grant-in-Aid for Scientific Research (17540275, 20244035) of the Ministry of Education, 
Science, Sports and Culture of Japan, and the JSPS Core-to-Core 
Program, International Research Network for Exotic Femto Systems (EFES). 
MK acknowledges the support by Grant-in-Aid for JSPS Fellows (18.11384). 
\end{acknowledgments}



\end{document}